\documentclass[twocolumn,preprintnumbers,superscriptaddress,amsmath,amssymb,aps,pre,10pt]{revtex4-1}

\usepackage{graphicx}
\usepackage{dcolumn}
\usepackage{bm}
\usepackage{color}
\usepackage{multirow}
\usepackage{listings}

\begin{document}

\title{Systemic risk and spatiotemporal dynamics of the US housing market}

\author{Hao Meng}
 \affiliation{School of Business, East China University of Science and Technology, Shanghai 200237, China} %
 \affiliation{School of Science, East China University of Science and Technology, Shanghai 200237, China} %

\author{Wen-Jie Xie}
 \affiliation{School of Business, East China University of Science and Technology, Shanghai 200237, China} %
 \affiliation{School of Science, East China University of Science and Technology, Shanghai 200237, China} %

\author{Zhi-Qiang Jiang}
 \affiliation{School of Business, East China University of Science and Technology, Shanghai 200237, China} %
 \affiliation{Research Center for Econophysics, East China University of Science and Technology, Shanghai 200237, China} %

\author{Boris Podobnik}
 \affiliation{Department of Physics and Center for Polymer Studies, Boston University, MA, USA} %
 \affiliation{Zagreb School of Economics and Management, 10000 Zagreb, Croatia} %
 \affiliation{Faculty of Civil Engineering, University of Rijeka, 51000 Rijeka, Croatia} %

\author{Wei-Xing Zhou}
 \email{wxzhou@ecust.edu.cn}
 \affiliation{School of Business, East China University of Science and Technology, Shanghai 200237, China} %
 \affiliation{School of Science, East China University of Science and Technology, Shanghai 200237, China} %
 \affiliation{Research Center for Econophysics, East China University of Science and Technology, Shanghai 200237, China} %

\author{H. Eugene Stanley}
\email{hes@bu.edu}
\affiliation{Department of Physics and Center for Polymer Studies, Boston University, MA, USA}

\date{\today}

\begin{abstract}
  Housing markets play a crucial role in economies and the collapse of a real-estate bubble usually destabilizes the financial system and causes economic recessions. Since the recent global financial tsunami and follow-up economic crisis triggered by the US subprime mortgage crisis, there is increasing interest in the investigation of the ripple effect and cross-sectional convergence of house prices of different regions. However, the complex evolving behavior of the housing market is still not well understood. We investigate the systemic risk and spatiotemporal dynamics of the US housing market (1975/Q1 to 2011/Q4) at the state level. We find that the largest eigenvalue $\lambda_1$ of the correlation matrix which commonly reflects a common market effect has an upward trend roughly since 1993 and experienced additional boost around 2008 that is again in agreement with the bursting of the housing bubble and financial crisis of 2007-2010. We surprisingly find there are time periods during which the market effect of $\lambda_1$ is weak and shows a partitioning ability, while other deviating eigenvalues exhibit a weak market effect. We also unveil that the component signs of the eigenvectors contain either geographical information or the extent of differences in house price growth rates or both. According to the information content embedded in the largest eigenvalues, we show that the US housing market experienced six regimes, which is consistent with the evolution of state clusters identified by performing the box clustering algorithm and the consensus clustering algorithm on the partial correlation matrices. In the early regimes, only a relative small number of states formed clusters, whose constituents vary remarkably. In the late regimes, the clusters were quite stable within each regime, and split and merged when the market transitioned from one regime to another. Our analysis uncovers that dramatic increases in the systemic risk are usually accompanied with regime shifts, which provides a means of early detection of housing bubbles.
\end{abstract}


\maketitle

Because houses and apartments are tradable and are commonly used in speculations, they are considered as a special kind of commodity. As time passes the house prices boom and bust. Because the housing market is closely related to the financial system and plays a crucial role in economies, a crash of the housing market usually has disastrous consequences, causing financial crisis and economic recession. Recent examples include the 1997--1998 Asian crisis \cite{Kaminsky-Reinhart-1999-AER,Quigley-2001-JHE,Fung-Forrest-2002-HS} and the 2007--2012 global financial tsunami followed by the 2008--2012 global recession and the European sovereign-debt crisis, none of which has ended  \cite{Sanders-2008-JHE}. When the correlations among the constituents of a market become stronger and the ripple effect increases \cite{Giussani-Hadjimatheou-1991-PRS}, prices tend to converge \cite{Kim-Rous-2012-JHE} and the systemic risk increases. However, there is evidence showing that alternative measures based eigenvalues and eigenvectors of correlation matrix outperform the average correlation in quantifying systemic risks, characterizing market integration and constructing profitable investment portfolios \cite{Pukthuanthong-Roll-2009-JFE,Billio-Getmansky-Lo-Pelizzon-2012-JFE,Kritzman-Li-Page-Rigobon-2011-JPM}. Hence, it is extremely important to understand the spatiotemporal dynamics of housing markets through an investigation of the correlation matrix of price growth rates.

The correlation matrices of stock returns and indices have been widely studied in different markets \cite{Tumminello-Lillo-Mantegna-2010-JEBO}. The studies have employed variety of methods ranging from the minimal spanning trees \cite{Mantegna-1999-EPJB}, the planar maximally filtered graph \cite{Tumminello-Aste-DiMatteo-Mantegna-2005-PNAS} based on distance matrices, to RMT \cite{Laloux-Cizean-Bouchaud-Potters-1999-PRL,Plerou-Gopikrishnan-Rosenow-Amaral-Stanley-1999-PRL}. All methods can be used to identify constituent clusters in financial systems \cite{Tumminello-Lillo-Mantegna-2010-JEBO}. When RMT is applied to investigate the correlation structure of financial markets, the largest eigenvalue serves to explain the collective behavior of the market, and other eigenvalues are commonly used to explain clustering of stocks or indices into groups with specific traits.

The correlation matrices of housing markets are  rarely studied, mainly due to the short length of house price indices, where the sampling frequency is usually either monthly or quarterly. In this work, within the RMT framework at the state level, we investigate the spatiotemporal dynamics of the US housing market. We analyze the All-Transactions Indexes of the 50 states including the District of Columbia, which are estimated with sales prices and appraisal data and published by the Federal Housing Finance Agency. The data are quarterly recorded beginning at 1975/Q1 and ending at 2011/Q4 giving in total 148 quarters.

We denote $S_i(t)$ the quarterly housing price index (HPI) of US state $i$ at time
$t$. The logarithmic return at time $t$ is defined as
\begin{equation}
  r_i(t) = \ln S_i(t) - \ln S_i(t-1).
  \label{vector}
\end{equation}
For each moving window $[t-s+1,t]$ at time $t$ of size $s$, we compute the correlation matrix $\mathbf{C}(t)$, whose element $C_{ij}$ is the Pearson correlation coefficient between the return time series of US states $i$ and $j$,
\begin{equation}
  C_{ij}(t)=\frac{1}{\sigma_i
    \sigma_j}\sum_{k=t-s+1}^t\left[r_i(k)-\mu_i][r_j(k)-\mu_j\right],
 \label{C}
\end{equation}
where $\mu_i$ and $\mu_j$ are the sample means and $\sigma_i$ and
$\sigma_j$ are the standard deviations
 of the two
states $i$ and $j$ respectively.

It is known that stock markets are characterized by both fast and slow dynamics \cite{Drozdz-Grummer-Gorski-Ruf-Speth-2000-PA,Song-Tumminello-Zhou-Mantegna-2011-PRE}.In order to estimate the empirical correlation matrix and minimize the unavoidable statistical uncertainty, we need to use a large window containing a large number of data points. On the other hand, large windows reduce our ability to investigate the fast dynamics in correlation studies. In addition, the correlation matrix is no longer invertible \cite{Billio-Getmansky-Lo-Pelizzon-2012-JFE,Song-Tumminello-Zhou-Mantegna-2011-PRE} when $s$ is smaller than the number of time series that is in our study $51$ (the number of states), implying $s_{\min}=51$. Here, we choose $s=60$ quarters, which gives $89$ moving windows for investigation.


\vspace{5mm}
\noindent{\Large\textbf{Results}}
\vspace{3mm}

\begin{figure}[htbp]
  \centering
  \includegraphics[width=7.5cm]{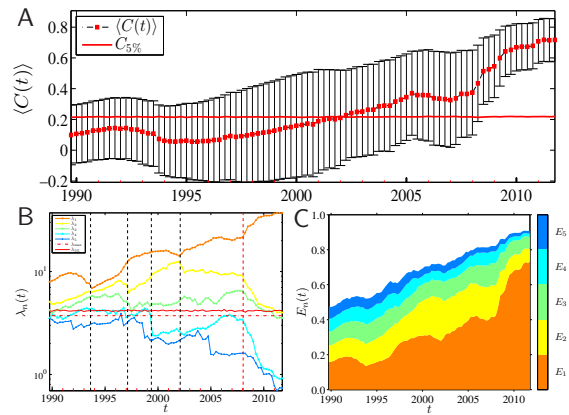}
  \caption{\label{Fig:RMT:HPI:lambda:1234} ({\it{A}}) Evolution of the average correlation coefficient. The horizontal red line shows the critical value at significance level 5\% of the correlation coefficient at each time $t$. The error bar is the standard deviation of the PDF at each time $t$. For the evolution of the PDF, see {\color{blue}{Fig.~S1}}. ({\it{B}}) Evolution of the five largest eigenvalues $\lambda_n$ of $\mathbf{C}(t)$ with $n=1,2,3,4$, and $5$. The horizontal dot-dashed red line is the maximum eigenvalue $\lambda_{\max}$ predicted by the RMT and the horizontal red line represents the critical values $\lambda_{5\%}$ at the significance level of 5\%. The five vertical dashed lines corresponding to the five regime-shift points. ({\it{C}}) Evolution of absorption ratio $E_n(t)$ for $n=1,2,3,4$, and $5$.} 
\end{figure}

\noindent{\textbf{Correlation coefficient.}}
In Fig.~\ref{Fig:RMT:HPI:lambda:1234}{\it{A}}, we show the average correlation coefficient of Eq.~{\bf{\ref{C}}} calculated for each year during the last two decades. In recent years the average correlation coefficient has substantially increased implying that the US  housing market has become strongly correlated. In prior years of the period studied, we find that only for a small number of states their housing indexes were correlated that is in contrast with the past decade where we find a sharp increase in housing market correlations, which implies a sharp increase in systemic market risk.

\noindent{\textbf{Eigenvalues.}}
For each $t$ larger or equal to $t = 1990/Q1$, we calculate the correlation matrix $\mathbf{C}(t)$ and  compute its 51 eigenvalues $\{\lambda_n: n=1,\cdots,51\}$. Then we sort the eigenvalues $\{\lambda_n\}$ in the descending order, and calculate the corresponding eigenvectors $\mathbf{u}_n(t)=[u_{n,1}(t),\cdots,u_{n,51}(t)]^{\mathrm{T}}$.

If $\mathbf{M}$ is a $T\times N$ matrix with mean 0 and variance $\sigma^2=1$, one defines $\mathbf{C}=\frac{1}{T}\mathbf{M}^{\mathrm{T}}\mathbf{M}$. In the limit $N\rightarrow \infty$, $T\rightarrow \infty$ where $Q=T/N\geq1$ is fixed, the probability density $f_{\mathrm{RMT}}(\lambda)$ of eigenvalues $\lambda$ of matrix $\mathbf{C}$ is $f_{\mathrm{RMT}}(\lambda)=\frac{Q}{2\pi}{\sqrt{(\lambda_{\max}-\lambda)(\lambda-\lambda_{\min})}}/{\lambda}$, where $\lambda\in[\lambda_{\min}, \lambda_{\max}]$ and $\lambda_{\min,\max} = 1+1/Q\pm2\sqrt{1/Q}$ \cite{Sengupa-Mitra-1999-PRE,Laloux-Cizean-Bouchaud-Potters-1999-PRL,Plerou-Gopikrishnan-Rosenow-Amaral-Guhr-Stanley-2002-PRE}. If an eigenvalue $\lambda$ is greater than $\lambda_{\max}$---so deviating from the prediction of the RMT---its eigenvector usually contains valuable information about the market dynamics. However, for real data, the limit conditions $N\rightarrow\infty$ and $T\rightarrow \infty$ are never fulfilled and some finite-size effect should be included in the RMT studies. To this end, in order to identify the deviating eigenvalues, we randomize the housing indexes time series to destroy any temporal correlations. Then we calculate a new correlation matrix $\mathbf{C}_{\mathrm{Rnd}}$ from the randomized return time series, and compute corresponding 51 eigenvalues. Repeating this procedure 1000 times we obtain in total 51,000 eigenvalues based on which we calculate the probability density of eigenvalues $f_{\mathrm{Rnd}}(\lambda)$. Although the density functions $f_{\mathrm{RMT}}(\lambda)$ and $f_{\mathrm{Rnd}}(\lambda)$ overlap to a great degree, they exhibit some differences in the right tail. We find that $f_{\mathrm{Rnd}}(\lambda)$ is not bounded by the maximum eigenvalue $\lambda_{\max}$ predicted by the RMT ({\color{blue}{Fig.~S2}}), which is due to the fact that the HPI returns have fat tails.

The question of whether housing and generally financial bubbles can be identified in advance is one of the topics in economics theory. To this end, it is observed in Fig.~\ref{Fig:RMT:HPI:lambda:1234}{\it{B}} that the largest eigenvalue $\lambda_1$ of $\mathbf{C}(t)$ has an upward trend  roughly since 1993. We note that $\lambda_1$ experienced additional boost around 2008 that is again in agreement with the bursting of the real estate bubble and the financial crisis of 2007-2010 around the world. Fig.~\ref{Fig:RMT:HPI:lambda:1234}{\it{B}} illustrates that the largest eigenvalue $\lambda_1$ of $\mathbf{C}(t)$ is larger than the maximum eigenvalue $\lambda_{\max}$ predicted by the RMT and also larger than the critical value $\lambda_{5\%}$ of $f_{\mathrm{Rnd}}(\lambda)$. For the second largest eigenvalue, we find $\lambda_2 >  \lambda_{\max}$ for all $\mathbf{C}(t)$ matrixes and $ \lambda_2 >  \lambda_{5\%}$ for most $\mathbf{C}(t)$ matrixes. We also find that the third largest eigenvalue $\lambda_3$ is larger than $\lambda_{\max}$ and $\lambda_{5\%}$ for most $\mathbf{C}(t)$ matrixes, and the fourth largest eigenvalue $\lambda_4$ is larger than $\lambda_{\max}$ and $\lambda_{5\%}$ for part of the $\mathbf{C}(t)$ matrixes. In contrast, the fifth largest eigenvalue $\lambda_5$ falls well within the bulks of $f_{\mathrm{RMT}}(\lambda)$ and $f_{\mathrm{Rnd}}(\lambda)$ ({\color{blue}{Fig.~S2}}). Therefore, the eigenvalues $\lambda_1$, $\lambda_2$ and $\lambda_3$ should contain information about nontrivial spatiotemporal properties of the US housing market dynamics. We also include $\lambda_4$ in our investigation.

A better measure of systemic risk is the absorption ratio $E_n=\sum_{i=1}^n\lambda_i/N$ \cite{Pukthuanthong-Roll-2009-JFE,Billio-Getmansky-Lo-Pelizzon-2012-JFE,Kritzman-Li-Page-Rigobon-2011-JPM}, which is shown in Fig.~\ref{Fig:RMT:HPI:lambda:1234}{\it{C}}. It is observed that the systemic risk increases almost linearly, even after the recent housing bubble bust in 2007, which indicates that the US housing market becomes more instable continually and the market is very fragile at high value.

\noindent{\textbf{Collective market effect and regime shifts.}}
For each eigenvalue $\lambda_n$, we can construct its eigenportfolio, whose returns are calculated by
\begin{equation}
  R_n(t') = \mathbf{u}_n^{\mathrm{T}}(t') \cdot \mathbf{r}(t')
\end{equation}
where $t'=t-s+1,\cdots,t$, and $\mathbf{r}(t') =[r_{1}(t'),\cdots,r_{51}(t')]^{\mathrm{T}}$ is a vector whose components are state-level HPI returns defined in Eq.~{\bf\ref{vector}}. To evaluate the collective market effect embedded in $\lambda_n$, we investigate the following linear regressive model between $R_n(t')$ and the return $R(t')$ of the US HPI
\begin{equation}
  R_n(t') = k_n(t) R(t') +\epsilon(t'),
\end{equation}
where $R_n$ and $R$ are normalized respectively to zero mean and unit variance \cite{Plerou-Gopikrishnan-Rosenow-Amaral-Guhr-Stanley-2002-PRE}, and $k_n(t)$ is the correlation coefficient between $R_n$ and $R$ in time $t'$. To estimate the value of $k_n$, we perform  an ordinary least-squares (OLS) linear regression together with a robust regression. Since the results and conclusions for both methods are qualitatively virtually the same, in the following we discuss the OLS results only. Before we proceed with the results for the housing market, we note that for stock markets, it is obtained that $k_1$ is significantly different from 0 and is usually close to 1, while $k_n\approx0$ for $n>1$ \cite{Plerou-Gopikrishnan-Rosenow-Amaral-Guhr-Stanley-2002-PRE}. In other words, for the stock market the largest eigenvalue reflects the common behavior of the market, while the rest of the eigenvalues do not contain information about such a market effect.

\begin{figure}[htb]
  \centering
  \includegraphics[width=8.4cm]{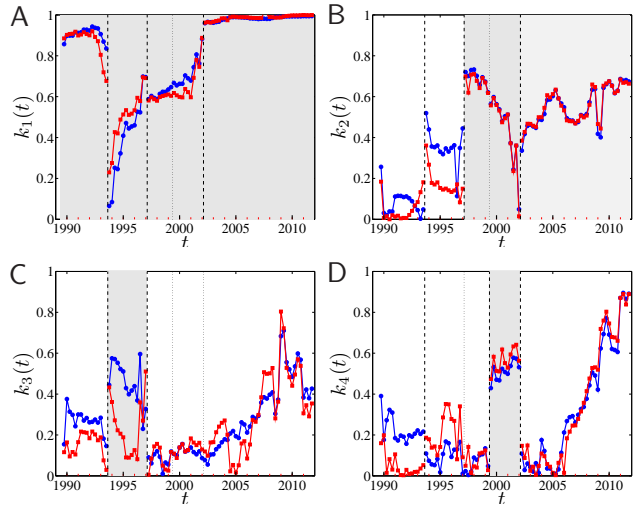}
  \caption{\label{Fig:RMT:HPI:MarketEffect} Market effect hidden in the largest eigenvalues. ({\em{A}} to {\em{D}}) Evolution of the correlation coefficient $k_n(t)$ between $R_n$ and $R$ in each moving window. The blue symbols are estimated using ordinary least-squares linear regression, while the red ones are estimated using robust fitting. The four vertical lines indicate four regime-shift points ${\mathcal{T}}_1$ between $1993Q3$ and 1993Q4, ${\mathcal{T}}_2$ between $1997Q1$ and 1997Q2, ${\mathcal{T}}_3$ between $1999Q2$ and 1999Q3, and ${\mathcal{T}}_4$ between $2002Q2$ and 2002Q3, separating five different regimes. The shading area in each plot means that the associated eigenvalue contains a market effect in the corresponding time period. See {\color{blue}{Fig.~S3}} for the scatter plots of $R_n$ against $R$.
  }
\end{figure}

In the following, we report that the RMT results obtained for the US housing market substantially differ from the results obtained for stock markets (Fig.~\ref{Fig:RMT:HPI:MarketEffect} and {\color{blue}{Fig.~S3}}). For the housing market, we observe that the correlation coefficient $k_1$ between $R(t')$ and $R_1(t')$ is large for the first four years, and then experience a sudden drop from 0.8354 ($1993Q3$) to 0.0655 ($1993Q4$). Then we find that $\lambda_1$ increase gradually to 0.8826 ($2002Q2$) and 0.9593 ($2002Q3$) and then remain at a high level close to 1. This behavior for $\lambda_1$ over time indicates that we can approximately identify three regimes for three time periods: $[1989Q4, 1993Q3]$, $[1993Q4, 2002Q2]$ and $[2002Q3, 2011Q4]$. We surprisingly reveal that the two regime-shift points in Fig.~\ref{Fig:RMT:HPI:MarketEffect} virtually overlap with the first two local minima in the time dependence of $\lambda_1$ in Fig.~\ref{Fig:RMT:HPI:lambda:1234}. Therefore, in the regimes corresponding to the first and last time periods, the market effect quantified by the correlation coefficient $k_1$ is remarkable; In contrast, the market effect is much weaker in the second time period ({\color{blue}{Fig.~S3}}). Within the second time period, we further identify a regime-shift point between $1997Q1$ and $1997Q2$, where $k_1$ drops from 0.6955 to 0.5879.

For the second largest eigenvalue $\lambda_2$, we identify three regime shifts: $1993Q3$--$1993Q4$, $1997Q1$--$1997Q2$, and $2002Q2$--$2002Q3$ (Fig.~\ref{Fig:RMT:HPI:lambda:1234}{\textit{B}}). Surprisingly, these regime-shift points are identical to those we found for $\lambda_1$. For the third largest eigenvalue $\lambda_3$, we identify two regime shifts: $1993Q3$--$1993Q4$ and $1997Q1$--$1997Q2$ that correspond to first and second regime shifts we found in eigenvalues $\lambda_1$ and $\lambda_2$ (Fig.~\ref{Fig:RMT:HPI:lambda:1234}{\textit{C}}). Finally, for the fourth largest  eigenvalue $\lambda_4$, we identify three regime shifts: $1993Q3$--$1993Q4$, $1999Q2$--$1999Q3$, and $2002Q2$--$2002Q3$ where first and third regime shifts  correspond to those we found for the eigenvalues $\lambda_1$ and $\lambda_2$ (Fig.~\ref{Fig:RMT:HPI:lambda:1234}{\textit{D}}). Based on four different regime shifts in Figs.~\ref{Fig:RMT:HPI:lambda:1234}{\textit{A}} to {\textit{D}}, we identify five regimes in the eigenvalues: ${\mathcal{R}}_1=[1989Q4, 1993Q3]$, ${\mathcal{R}}_2=[1993Q4, 1997Q1]$, ${\mathcal{R}}_3=[1997Q2, 1999Q2]$, ${\mathcal{R}}_4=[1999Q3, 2002Q2]$, and ${\mathcal{R}}_5=[2002Q3, 2011Q4]$, revealing an interesting dynamics on the US housing market.

We find that in regime ${\mathcal{R}}_1$, only for the largest eigenvalue $\lambda_1$ the market effect---quantified by the correlation coefficient $k_1$ between $R(t')$ and $R_1(t')$--- is substantially large ({\color{blue}{Fig.~S3}}). In regime ${\mathcal{R}}_2$, the market effect for $\lambda_1$ becomes substantially weaker than in regime ${\mathcal{R}}_1$, and $\lambda_3$ exhibits a moderately stronger market effect only at some time $t$ ({\color{blue}{Fig.~S3}}). In regime ${\mathcal{R}}_3$, $\lambda_1$ and $\lambda_2$ exhibit a substantially stronger market effect than $\lambda_3$ and $\lambda_4$. In regime ${\mathcal{R}}_4$, $\lambda_1$, $\lambda_2$, and $\lambda_4$ exhibit a strong market effect. Finally, in regime ${\mathcal{R}}_5$, only $\lambda_1$ exhibits a strong market effect, while the rest of eigenvalues $\lambda_2-\lambda_4$ do not. Therefore, we find that the largest eigenvalue $\lambda_1$ almost always exhibits a market effect, whereas the other eigenvalues exhibit a market effect only frequently especially when the market effect becomes weak in $\lambda_1$.

\begin{figure}[htb]
\centering
\includegraphics[width=8.4cm]{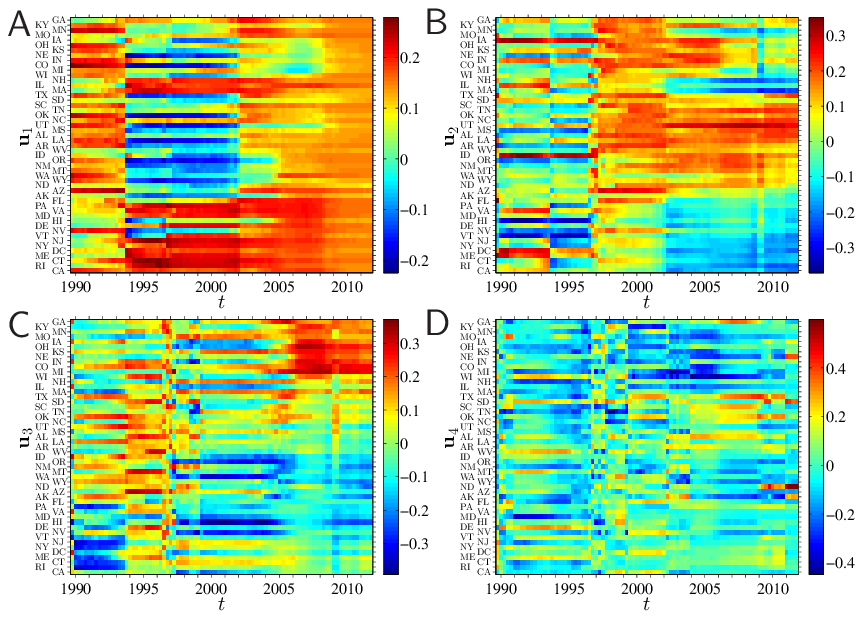}
\caption{\label{Fig:RMT:HPI:Eigenvectors} Evolution of the eigenvectors
  of the largest eigenvalues: ({\textit{A}}) $\mathbf{u}_1$,
  ({\textit{B}}) $\mathbf{u}_2$, ({\textit{C}}) $\mathbf{u}_3$, and
  ({\textit{D}}) $\mathbf{u}_4$. The five regimes ${\mathcal{R}}_1$ to
  ${\mathcal{R}}_5$ are visible. Moreover, we observe that the regime
  ${\mathcal{R}}_5$ can be separated into two regimes at $2007Q1$ to
  $2007Q2$ according to the evolution of $\mathbf{u}_3$.}
\end{figure}

\noindent{\textbf{Information contented in the eigenvectors associated with the largest eigenvalues.}}
For stock markets it was found that the components of the eigenvector of the largest eigenvalue are practically always positive where components exhibit small fluctuations over time, reflecting a market effect. The rest of the eigenvectors of other largest eigenvalues describe different clusters of stocks or industrial sectors \cite{Plerou-Gopikrishnan-Rosenow-Amaral-Guhr-Stanley-2002-PRE,Pan-Sinha-2007-PRE,Shen-Zheng-2009a-EPL}. For the US housing market, we find that the eigenvectors of the largest eigenvalues contain much richer information (Fig.~\ref{Fig:RMT:HPI:Eigenvectors} and {\color{blue}{Fig.~S4}}). The existence of five regimes ${\mathcal{R}}_1$ to ${\mathcal{R}}_5$ is vivid and the eigenvector components persist in each regime. Moreover, the graphical approach in Fig.~\ref{Fig:RMT:HPI:Eigenvectors} revealed that the regime ${\mathcal{R}}_5$ can be separated into two regimes at $2007Q1$ to $2007Q2$ according to the evolution of $\mathbf{u}_3$.

\begin{figure*}[tb]
  \centering
  \includegraphics[width=17cm]{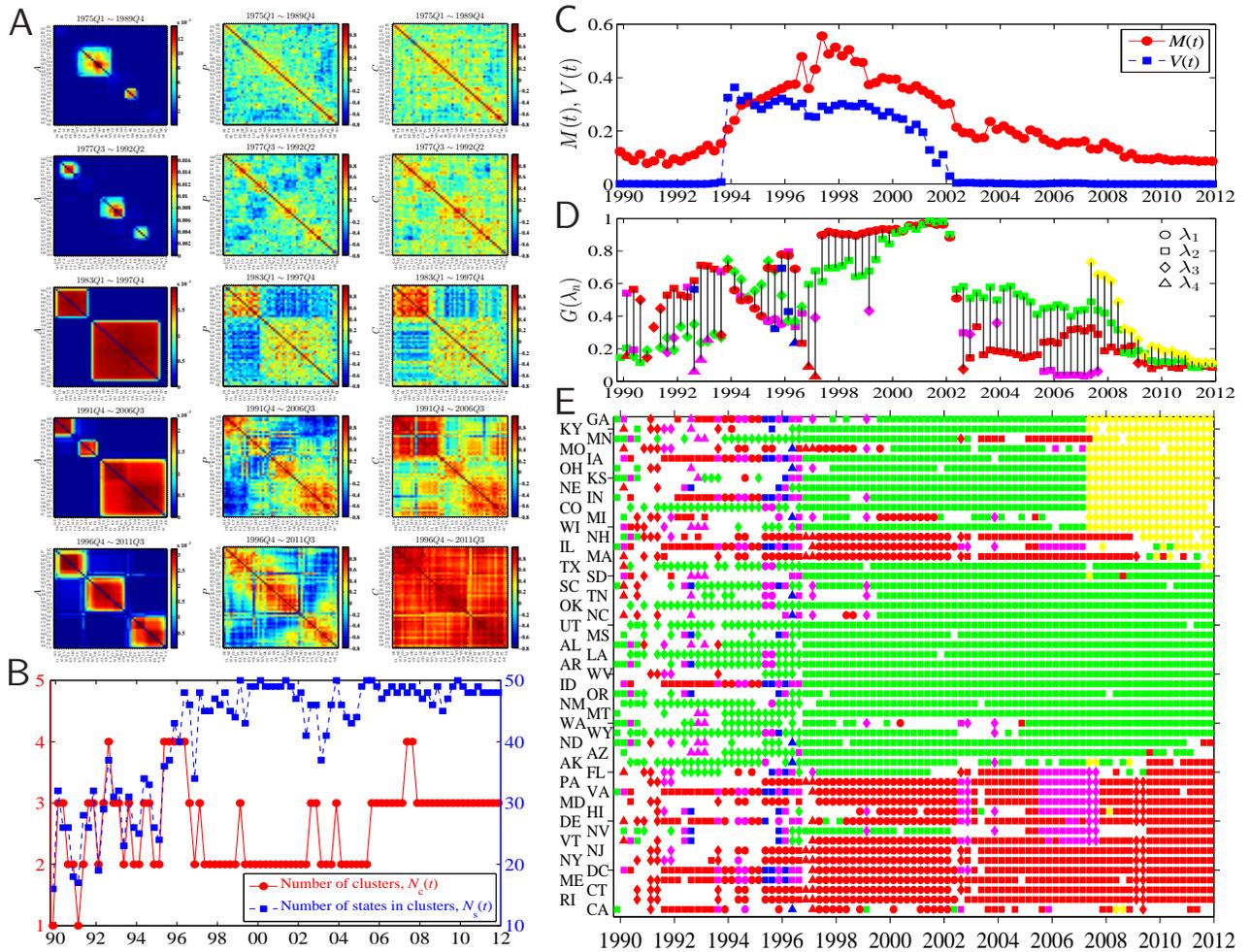}
  \caption{Evolution of the states clusters.  ({\textit{A}}) Typical affinity matrices ${\mathbf{A}}(t)$ (left column), partial correlation matrices ${\mathbf{P}}(t)$ (middle column), and    correlation matrices ${\mathbf{C}}(t)$ (right column). The order of the states is the same for the three matrices in each row. The ending quarters $t$ of the windows from top to bottom are $1989Q4$, $1992Q2$, $1997Q4$, $2006Q3$, and $2011Q3$.  ({\textit{B}}) Number of clusters $N_{\mathrm{c}}(t)$ and the corresponding number $N_{\mathrm{s}}(t)$ of states included in the detected clusters for each window.  ({\textit{C}}) Evolution of modularity $M(t)$ and the squared sum $V(t)$ of negative components in $\lambda_1(t)$. ({\textit{D}}) Maximal information ratio $G(\lambda_n)$ of certain eigenvalue $\lambda_n$ contributed to a cluster. Each cluster is represented by a colorful symbol. The determination of symbols and their coloring is explained in {\it{Methods}}.  ({\textit{E}}) Evolution of states clusters, where the order of the states is the same as ${\mathbf{A}}(t)$ at $t=2009Q3$. The states in a certain cluster are assigned with a cluster-specific colorful symbol and no symbol is assigned to those states not in any cluster. The colorful symbols have the same meaning as those in {\textit{D}}.}
  \label{Fig:RMT:HPI:EvolvingClusters}
\end{figure*}

Starting with the first eigenvector $\mathbf{u}_1$, we study its components over time, precisely for different regimes. We find that  in regime ${\mathcal{R}}_1$, almost all the components of $\mathbf{u}_1$ are positive. In contrast, after $1993Q4$ during the three regimes ${\mathcal{R}}_2$ to ${\mathcal{R}}_4$, many components of the first eigenvector $\mathbf{u}_1$ turn from positive to negative, as vividly  seen in Fig.~\ref{Fig:RMT:HPI:Eigenvectors}. During the period from $1993Q4$ to $2002Q2$, positive components of $\mathbf{u}_1$ mainly correspond to the states in the Eastern half of the US with exception of California and Arizona in the Western US. It means that the largest eigenvalue $\lambda_1$ partitions the US states into two groups. Since the states with positive components include predominantly the states with high HPI values, $\lambda_1$ still exhibits a modest market effect. As time passes  transferring from regime ${\mathcal{R}}_4$ to regime ${\mathcal{R}}_5$, states with initially negative components turn  from negative to positive components.

For the eigenvector $\mathbf{u}_2$, in the first two regimes ${\mathcal{R}}_1$ and ${\mathcal{R}}_2$ we find a comparable number of negligible positive and negative components, and it is not completely clear what information is contained in the US states with positive and negative components. Around $1997Q2$, the number of states with negative $\mathbf{u}_2$ components become significantly lower, and so the large majority of states have positive components {\it  reflecting a market effect}. The predomination of positive components over negative components persisted in ${\mathcal{R}}_3$ and ${\mathcal{R}}_4$. Since late ${\mathcal{R}}_4$, the $\mathbf{u}_2$ components of Washington and California  turn from positive to  negative, and then some Northeastern states do the same. In regime ${\mathcal{R}}_5$, the two clusters of states, one  with positive and the other with negative $\mathbf{u}_2$ components virtually correspond to states with low and high HPI growth rates, respectively, as identified by the super-exponential growth model \cite{Zhou-Sornette-2006b-PA}.

In the evolution of the eigenvector $\mathbf{u}_3$,  we find two interesting features. First, the majority of  states in regime ${\mathcal{R}}_2$ have positive components, reflecting a modest market effect. Second, there is an evident  subregime  around $2007Q2$, which surprisingly corresponds to the onset of the primary US mortgage crisis. The information contained in other regimes is ambiguous. Also, it is not easy to extract clear information from the evolution of the fourth eigenvector $\mathbf{u}_4$.

\noindent{\textbf{Evolution of state clusters.}}
To better understand the spatiotemporal dynamics of the US housing market at the state level, we partition the states into clusters for each time $t$. Because there is a strong market effect in the correlation matrices, the Pearson correlation coefficient between the return time series $r_i$ and $r_j$ of two US states $i$ and $j$ may not reflect their intrinsic relationship, but may be a reflection of a similar influence of the overall US HPI return $r_{\mathrm{us}}$ on $i$ and $j$ \cite{Kenett-Shapira-BenJacob-2009-JPS,Kenett-Tumminello-Madi-GurGershgoren-Mantegna-BenJacob-2010-PLoS1}. We thus utilize a clustering algorithm using the corresponding partial correlation matrices $\mathbf{P}(t)$ by removing the market effect. The partial correlation coefficient $P_{ij}$ between $r_i$ and $r_j$ with respect to $r_{\mathrm{us}}$ can be calculated as follows \cite{Baba-Shibata-Sibuya-2004-ANZJS,Kenett-Tumminello-Madi-GurGershgoren-Mantegna-BenJacob-2010-PLoS1}:
\begin{equation}
  P_{ij} = \frac{C_{ij}-C_{i,\mathrm{us}}C_{j,\mathrm{us}}} {\sqrt{\left(1-C_{i,\mathrm{us}}^2\right)\left(1-C_{j,\mathrm{us}}^2\right)}},
\end{equation}
where $C_{i,\mathrm{us}}$ ($C_{j,\mathrm{us}}$) is the Pearson
correlation coefficient between $r_i$ ($r_j$) and $r_{\mathrm{us}}$.

For each partial correlation matrix $\mathbf{P}(t)$, we combine the box clustering and consensus clustering methods to search for clusters of states \cite{SalesPardo-Guimera-Moreira-Amaral-2007-PNAS,Lancichinetti-Fortunato-2012-SR}. We first determine the optimal ordering of $\mathbf{P}(t)$ by identifying the largest elements in $\mathbf{P}(t)$ closest to the diagonal, where the simulated annealing approach is adopted to minimize the cost function
\begin{equation}
  Q = \sum_{i,j=1}^{51}|i-j|P_{ij}(t).
\end{equation}
We then use a greedy algorithm to partition clusters of states and isolated states \cite{SalesPardo-Guimera-Moreira-Amaral-2007-PNAS}. This procedure we repeat 200 times and we obtain 200 partitions. We construct an affinity matrix ${\mathbf{A}}'$, whose  element $A_{ij}'$ is the number of partitions in which $i$ and $j$ were assigned to the same cluster, divided by the number of partitions 200. Finally we apply the clustering method to the affinity matrix ${\mathbf{A}}'$, resulting in a final partition ${\mathbf{A}}(t)$ \cite{Lancichinetti-Fortunato-2012-SR}. For each $t$, we rearrange the order of states in ${\mathbf{P}}(t)$ and ${\mathbf{C}}(t)$ to be the same as in ${\mathbf{A}}(t)$.

The evolution of the three matrices is illustrated in {\color{blue}{Fig.~S5}}. In early years represented by regions ${\mathcal{R}}_1$ and ${\mathcal{R}}_2$, we  identify  the clusters of states (Fig.~\ref{Fig:RMT:HPI:EvolvingClusters}{\textit{A}}),  where the number of states forming clusters is relatively small (Fig.~\ref{Fig:RMT:HPI:EvolvingClusters}{\textit{B}}) and the constituent states of the clusters are unstable ({\color{blue}{Fig.~S5}})---these properties are consistent with the fact that the average cross-correlation level among US states is very low, indicating that the housing markets of different US states are to some extent isolated. With the development of the US housing market during the period $1996Q4$--$2002Q1$, more US states enter two different clusters of significantly different sizes (Fig.~\ref{Fig:RMT:HPI:EvolvingClusters}{\textit{B}}). This period roughly corresponds to the two regimes ${\mathcal{R}}_3$ and ${\mathcal{R}}_4$. During this period, we find that both clusters exhibit  relative stability (Fig.~\ref{Fig:RMT:HPI:EvolvingClusters}{\textit{B}} and {\color{blue}{Fig.~S5}}). In regime ${\mathcal{R}}_5$, we find that the smaller cluster further splits into two smaller clusters whereas the cluster  remain  relatively stable. Around $2007Q2$, the larger cluster splits into two clusters of comparable size, but shortly the two smaller clusters merged back into one cluster ({\color{blue}{Fig.~S5}}). Finally we find  three stable clusters of similar size, forming the sixth regime ${\mathcal{R}}_6$.

For each window $t$, there are up to four clusters of states and the number of states in each cluster varies from one window to another. For each cluster, one of the four deviating eigenvalues  has a dominant contribution (Fig.~\ref{Fig:RMT:HPI:EvolvingClusters}{\textit{D}}). We find that in regimes ${\mathcal{R}}_2$ and ${\mathcal{R}}_3$, the largest eigenvalue $\lambda_1$ participates in the cluster partitioning.

The spatiotemporal dynamics of the clusters of states is presented in Fig.~\ref{Fig:RMT:HPI:EvolvingClusters}{\textit{E}}. Roughly speaking, the states in the red cluster have larger price fluctuations (especially greater price value), while the states in the green cluster have smaller HPI growth rate fluctuations ({\color{blue}{Fig.~S6}}). In earlier years (${\mathcal{R}}_1$ and ${\mathcal{R}}_2$), the clusters are unstable with a large number of states shifting between clusters. During this period, the primary contribution to the green cluster comes from the third largest eigenvalue $\lambda_3$. In contrast, there are more eigenvalues contributing to the red cluster. In 1997 we find that two large stable clusters are rapidly formed in which the green and red clusters are dominated by $\lambda_2$ and $\lambda_1$, respectively. This phase-transition-like phenomenon in 1997 may have been a sign of a fast ripple effect within the US housing market. After $2005Q2$, the red cluster splits into two smaller clusters for approximately two years and almost all of the clusters are dominated by $\lambda_2$. We find that since $2007Q2$, the green cluster has partitioned into two smaller clusters: the red and green clusters are still dominated by $\lambda_2$ and the new yellow cluster is dominated by $\lambda_3$. The time period of these two transitions corresponds to the downturn in the US housing market. In a nutshell, Fig.~\ref{Fig:RMT:HPI:EvolvingClusters}{\textit{E}} shows the extreme complexity of the spatiotemporal dynamics of the US housing market.

In order to have finer resolution in characterizing systemic risk, we divided the 51 time series into 6 clusters according to the clusters of states revealed in Fig.~\ref{Fig:RMT:HPI:EvolvingClusters}{\textit{E}}. We form a sample with 6 return time series, each being randomly chosen from a cluster. The eigenvalues of the correlation matrices of the sample with moving window of size 8 quarters are determined. We repeat this procedure for 50 times and average the corresponding eigenvalues. We find that the systemic risk increased sharply in early 1990's and dropped to a relatively low level in late 1990's ({\color{blue}{Fig.~S7}}). The absorption ratio started to increase dramatically in 2003 and remains at historical high. Different from results of the analysis of 14 metropolitan housing markets in the United States \cite{Kritzman-Li-Page-Rigobon-2011-JPM}, our analysis shows that the systemic risk is still at its historical high after the housing bubble peaked.

\vspace{5mm}
\noindent{\Large\textbf{Discussion}}
\vspace{3mm}

We have investigated the complex spatiotemporal dynamics of the US housing market at the state level from the perspective of random matrix theory. At a large timescale, evolution of the market can be separated into three time periods. During the first time period ($1989Q4$ to $1997Q1$), the market exhibits low correlation and the largest eigenvalue reflects a market effect, while the next three largest eigenvalues contain partitioning information. During the second time period ($1997Q2$ to $2002Q2$), the correlation among the states is still low and the market effect of the largest eigenvalue becomes weaker. We find that the largest eigenvalue contains partitioning information and that the deviating eigenvalues exhibit a weak market effect. During the last period, the largest eigenvalue exhibits a strong market effect and its partitioning function disappears, which corresponds to the fact that the market integration becomes much stronger and exhibits sharply increasing average correlations. During this period, the partitioning of the states is mainly caused by the second largest eigenvalue. After the subprime crisis, the third largest eigenvalue shows a partitioning function.

The unveiled regime shifts imply the abrupt increases in the systemic risk of the US housing market and imply that the housing bubble that burst in 2007 has accumulated as early as 1997. Our finding is consistent with and provides convincing evidence for the conclusion based on the evolution of the absorption ratio \cite{Kritzman-Li-Page-Rigobon-2011-JPM}.

We observe that there are both positive and negative components in the eigenvectors of the deviating eigenvalues for most time windows. When the components of an eigenvector has the same sign, it usually reflects a market effect. When an eigenvector has both positive and negative components, especially when their amounts are comparable, the eigenvector may reflect either geographical information or differences in house price growth rates or both. The information contained in the signs of the eigenvector components has recently been reported for stock markets \cite{Yan-Liu-Zhu-Chen-2012-CPL,Jiang-Zheng-2012-EPL,Junior-2011-XXX}. However, the US housing market seems more complex than stock markets.

During the evolution of the US housing market, we observe that prices diffuse in significant and complex ways that do not require geographical bonds \cite{Pollakowski-Ray-1997-JHR}. The splitting and merging of clusters indicate that there is no national convergence of house prices. Furthermore, the model in Ref.~\cite{Kim-Rous-2012-JHE} in which there are several clusters within which the prices converge is too simple, although we have used a different approach for state clustering. It is rational to conjecture that there are different classifications for converging clusters in different time periods.

\vspace{5mm}
\noindent{\Large\textbf{Methods}}
\vspace{3mm}

\noindent{\textbf{Determination of symbols in Fig.~5{\normalsize\it{D}}.}}
We determine the symbol of each cluster according to the contribution of the eigenvalues. Note that the correlation matrix $\mathbf{C}(t)$ can be decomposed as \cite{Noh-2000-PRE,Kim-Jeong-2005-PRE}
\begin{equation}
  \mathbf{C}(t) = \sum_{n=1}^{51} \mathbf{C}_{\lambda_n}(t) = \sum_{n=1}^{51} \lambda_n(t) \mathbf{u}_n(t)\mathbf{u}_n^{\mathrm{T}}(t),
\end{equation}
where $\mathbf{C}_{\lambda_n}(t) = \lambda_n(t) \mathbf{u}_n(t)\mathbf{u}_n^{\mathrm{T}}(t)$ is the matrix associated with $\lambda_n$, and its element is $C_{\lambda_n,ij}(t) = \lambda_n(t) u_{n,i}(t)u_{n,j}(t)$. We define the information ratio of $\lambda_n$ in a certain cluster $\mathcal{C}(t)$ as
\begin{equation}
  G(\lambda_n,\mathcal{C}(t)) =
  \frac{\sum_{i,j\in\mathcal{C}(t)}C_{\lambda_n,ij}(t)}
       {\sum_{i,j\in\mathcal{C}(t)}C_{ij}(t)},
\end{equation}
which is the relative contribution of $\lambda_n$ to $\mathcal{C}(t)$, and the maximum information ratio $G(\lambda_n)$ can be simply determined. Since almost all the components of $\mathrm{u}_1$ are
positive in regimes ${\mathcal{R}}_1$, ${\mathcal{R}}_5$, and ${\mathcal{R}}_6$ (Fig.~\ref{Fig:RMT:HPI:EvolvingClusters}{\textit{C}}), the partitioning function of $\lambda_1$ is weak. In these time periods, the modularity defined in Ref.~\cite{Newman-Girvan-2004-PRE,Guimera-SalesPardo-Amaral-2004-PRE} is also relatively small. We thus exclude $\lambda_1$ from the determination of $G(\lambda_n)$ in these three regimes. If $\lambda_n(t)$ makes the largest contribution to cluster $\mathcal{C}(t)$ ({\em{i.e.}} $G(\lambda_n,\mathcal{C}(t))$ is maximal), then an eigenvalue-specific symbol is assigned to $\mathcal{C}(t)$: circle ({\large$\bullet$}) for $\lambda_1$, square ({\footnotesize$\blacksquare$}) for $\lambda_2$, diamond ({\small$\blacklozenge$}) for $\lambda_3$, and triangle ($\blacktriangle$) for $\lambda_4$.

\noindent{\textbf{Coloring the states in Fig.~5{\normalsize\it{E}}.}}
For a given time $t$, states belonging to the same cluster are marked with the same color and states belonging to different clusters are marked with different colors. For simplicity, we define for each $t$ a color configuration vector $\Phi_t$, the elements of which correspond to the 51 states in a predetermined order. The elements of $\Phi_t$ corresponding to each cluster are assigned a unique positive integer and the remaining elements not belonging to a cluster are assigned zeros. For two configurations $\Phi_t$ and $\Phi_{t'}$, we define a measure of similarity $J$,
\begin{equation}
  J(\Phi_t,\Phi_{t'}) = \frac{|\Phi_{t}\cup \Phi_{t'}|} {51-\sum_{i=1}^{51} \delta_{0,\Phi_{t,i}\Phi_{t',i}}},
\end{equation}
where $\delta_{x,y}$ is the Kronecker delta function, which is equal to 1 if $x=y$, and 0 otherwise. The ultimate task of maximizing globally $\sum_{t'=1}^{50}\sum_{t=t'+1}^{51}J(\Phi_t,\Phi_{t'})$ is impossible since the number of the parameters is too large (Fig.~\ref{Fig:RMT:HPI:EvolvingClusters}{\textit{B}}).

To solve the coloring problem, we adopt a heuristic algorithm. We determine the colors of the clusters reversely from $2011Q4$ to $1989Q4$. We separate the time period into two intervals:
$I_1=[1989Q4,1996Q1]$ and $I_2=[1996Q2,2011Q4]$. For $t=2011Q4$, there are three clusters of states colored yellow, green, and red, respectively. When we start to determine $\Phi_t$ for a given $t\in I_2$, all $\Phi_{t'}$ with $t'>t$ have already been determined. The configuration $\Phi_t$ is determined by maximizing $F_2(\Phi_t) = \sum_{\tau=1Q}^{t'} J(\Phi_t,\Phi_{t+\tau})$, where $t'=\min\{6Q,2011Q4-t\}$. When $t\in I_1$, we maximize $F_1(\Phi_t) = \sum_{t'=1997Q1}^{1998Q3} J(\Phi_t,\Phi_{t'})$. Note that slightly varying the choice of the reference future configuration does not affect the results.



\vspace{5mm}
\noindent{\Large\textbf{Acknowledgements}}
\vspace{3mm}

\noindent{HM, WJX, ZQJ and WXZ received support from the National Natural Science Foundation of China Grant 11075054, the Shanghai (Follow-up) Rising Star Program Grant 11QH1400800, the Shanghai ``Chen Guang'' Project Grant 2012CG34, and Fundamental Research Funds for the Central Universities. BP and HES received support from the Defense Threat Reduction Agency (DTRA), the Office of Naval Research (ONR), and the National Science Foundation (NSF) Grant CMMI 1125290.}


\noindent{{\textbf{Supplementary information}} accompanies this paper is not available with this arXiv version because its size is too big. We will provide a link when the paper is published.}

\end{document}